\documentclass[11pt]{article}
\usepackage{erice,epsfig}

\begin{document}
\vspace*{4cm} \title{DIFFRACTION AT RHIC}

\author{A. BRAVAR$^a$, W. GURYN$^a$, 
S. R. KLEIN$^b$,
D. MILSTEAD$^c$,
B. SURROW$^a$}

\address{
$^a$ Brookhaven National Laboratory, Upton, NY, 11973, USA \break
$^b$ Lawrence Berkeley National Laboratory, Berkeley, CA, 94720, USA\break
$^c$ University of Liverpool, Liverpool, L69 3BX, UK\break
}
\vskip 0.1 in

\maketitle\abstracts{ The relativistic heavy ion collider (RHIC)
offers many opportunities to study diffraction in $pp$, $pA$ and $AA$
collisions.  Because both proton beams can be polarized, RHIC offers
the unique possibility of studying polarization effects in
diffraction.  We will introduce diffraction at RHIC and present three
compelling physics topics: hard diffraction with polarized beams,
identification of exotic mesons (non-$q\overline q$ states) in
double-Pomeron collisions, and using diffraction to measure the low-$x$
gluon density in $pA$ collisions, testing models of gluon saturation
and the colored glass condensate.  This note developed from discussion
at a workshop on ``Diffraction and Glueball Production at RHIC'' at
Brookhaven National Laboratory, May 17-18, 2002.}

\section{Introduction}

Diffractive events occur via the exchange of color singlet objects,
Pomerons, with the same quantum numbers as the vacuum ($J^{PC} =
0^{++}$).~\cite{nicolo} They are characterized by final states which
include rapidity gaps, regions of phase space containing no final
state particles.  Examples are $pp$ elastic scattering,
photoproduction of vector mesons, and diffractive $W^\pm$ production.
Typical hadronic interactions, in contrast, involve the exchange of
colored particles, where the probability of finding a particle-free
region with a width of $\Delta y$ units of rapidity is exponentially
suppressed as $\exp{(-\Delta ydN/dy)}$, where $dN/dy$ is the mean
number of particles per unit rapidity.  Meson exchange can also
produce rapidity gaps. However, the meson exchange contribution drops
with increasing collision energy, while the Pomeron exchange cross
section increases slightly.  So, at high (RHIC) energies, processes
with rapidity gaps are expected to be dominated by Pomeron exchange.

Despite 40 years of studying diffraction, many questions remain.  Soft
(low energy) diffraction is usually characterized in terms of the
optical model as represented by the absorptive part of the cross
section, and via a Regge trajectory.  The optical model can describe
phenomena such as elastic scattering and vector meson photoproduction,
but does very poorly with hard reactions like $W^\pm$ and jet
production.  These higher momentum-transfer reactions may be
explainable within perturbative QCD as via Pomeron
exchange,~\cite{forshaw} where the Pomeron is represented as a gluon
ladder (two gluons connected by additional gluon and quark loop
`rungs').  These models are moderately successful in explaining many
hard diffraction phenomena.

Despite these successes, many questions and controversies remain.  In
$pp$ collisions, it is still unclear whether diffractive jet
production can be explained as due to the collisions of two Pomerons.
Can the Pomeron flux from a relativistic proton be described in terms
of a Pomeron distribution, akin to the usual parton distributions?
Another example involves photoproduction.  At HERA, the $J/\psi$ and
$\psi'$ cross sections rise sharply with energy,~\cite{herajpsi} in
contrast to the expected slow energy dependence of the Pomeron
exchange model.  This has been argued to be a threshold effect, or due
to a hard component of the Pomeron.

The nature of the Pomeron is also debated.  In the simplest models, it
is almost entirely gluonic.  However, the observation of diffractive
deep inelastic scattering~\cite{dis} and diffractive $W^\pm$
production in $p\overline p$ collisions~\cite{wprod} both require a
significant quark component.  The role of the 3-gluon triangle
(Fig. 1b) is also debated.  It is also difficult to explain hard and
soft diffraction within a single theory;~\cite{landshoff} some papers
argue that they are two distinct phenomena.~\cite{landshoff2} For the
soft Pomeron, this is little understanding of how diffraction
(i.e. the absorptive part of the cross section) meshes with colored
QCD reactions.

RHIC can bring significant new data to bear on many of these
questions.  One key feature of RHIC is its polarized proton beams,
allowing for the first time studies of polarized hard diffraction up
to center of mass energies of 500 GeV.  As the first high energy $pp$
collider, RHIC can also look for differences between $pp$ and
$p\overline p$.

These studies will complement studies of $pp$ elastic scattering by
the $pp2pp$ collaboration. \cite{wlodek} The $pp2pp$ collaboration
will measure single and double spin asymmetries in elastic scattering,
and so probe the spin structure of the Pomeron at very soft momentum
scales.

Diffraction is also a tool to study meson spectroscopy.
Double-diffractive (two rapidity gaps) central production of mesons
has been studied extensively at the CERN SPS.~\cite{mesons} The
experiments found that invariant mass spectra changed drastically when
the protons were 'kicked' in opposite directions, compared to when the
protons went in the same direction.  This '$p_T$ filter' is still
poorly understood, but is clearly important for both meson
spectroscopy and for studies of diffraction.  RHIC studies of the spin
dependence of the $p_T$ filter could yield qualitatively new
information.

Events with rapidity gaps can also be a place to search for new
physics.~\cite{bj} The new physics can occur via either
double-Pomeron or two-photon interactions. The two event
classes look similar, except that in two-photon interactions, the
outgoing protons have a smaller average $p_T$; the classes may
be statistically separated on this basis.~\cite{pp}

Diffractive $pA$ collisions can be used to study the nucleus.  If the
Pomeron is made up of two gluons, then the Pomeron should exhibit
nuclear shadowing (the EMC effect), with a magnitude roughly the
square of the size of gluon shadowing.~\cite{dumitru}

Diffraction usually requires that the interacting protons remain
intact. Dedicated forward detectors called Roman pots are commonly
used to observe the outgoing protons.  Roman pots detect protons
travelling inside the beam pipe of the accelerator, making it possible
to measure very small scattering angles. Since the Roman pots are
usually placed past the accelerator magnets, they can act as
spectrometers, measuring the kinematics of the scattered protons. At
RHIC, diffraction could be studied by adding two Roman pot systems to
a central detector.  Because of the angular sensitivity of the $p_T$
filter and all spin measurements, pots with full azimuthal acceptance
are very desirable.

This physics was discussed at a workshop on ``Diffraction and Glueball
Searches at RHIC,'' at Brookhaven National Laboratory, May 17-18,
2002.~\cite{workshop} The workshop found ample
physics justification to install a Roman pot system around one of the
existing detectors. This writeup will now present three core topics
discussed at the workshop: meson spectroscopy, hard diffraction in
$pp$ and diffraction in $pA$.

\section{Meson Production in Soft $pp$ Diffraction}

\begin{figure}
\center{\psfig{figure=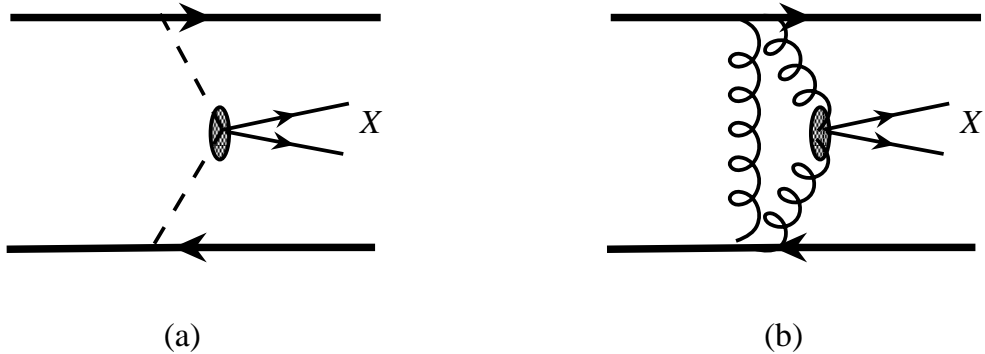,height=2.5in,clip=}}
\caption[]{The main diagrams for diffractive meson production.
(a) shows double Pomeron fusion, while (b) shows the gluon
triangle.}
\label{fig:feynman}
\end{figure}

Mesons can be produced diffractively via Pomeron-Pomeron fusion,
$pp\rightarrow ppPP\rightarrow ppX$, Figure~\ref{fig:feynman}.
Pomeron-meson and meson-meson fusion are also possible.  At lower
energies, the meson contributions are important; at RHIC, they should
be small.  Because Pomerons are believed to have a largely gluonic
content, double-Pomeron collisions may be a good channel to search for
glueballs.

Double-Pomeron production of mesons has been studied extensively by
experiments WA 76/91/102 at the CERN SPS.~\cite{mesons} These
experiments made a remarkable discovery: the character of the produced
mesons depended strongly on the direction of the momentum transfer
from the protons.  When the transverse momentum transfers pointed in
the same direction, one $K^+K^-$ (for example) invariant mass spectrum
was obtained, showing mostly conventional $q\overline q$ mesons.  When
the momentum transverse pointed in opposing directions, the $K^+K^-$
spectrum was very different, with many suspected 'exotic' (non
$q\overline q$) states appearing clearly.

\begin{figure}
\center{\psfig{figure=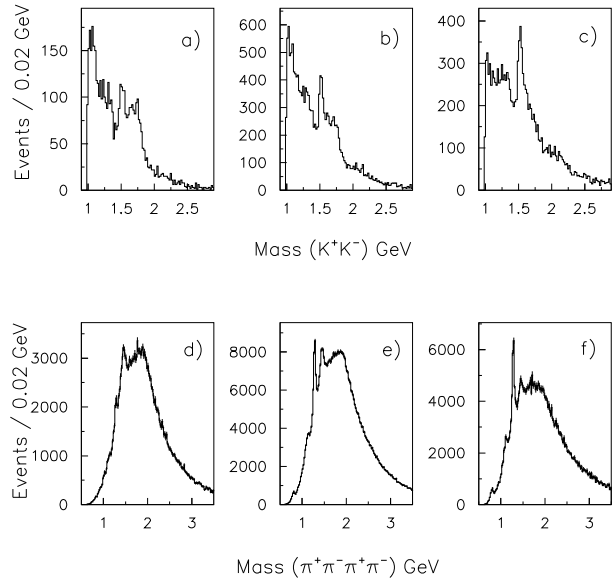,height=3.in,clip=}}
\caption[]{Invariant mass spectra for $K^+K^-$ final states for
(a)$dP_T<0.2$ GeV/c, (b)$0.2 {\rm GeV/c} dP_T<0.5$ GeV/c and
(c)$dP_T<0.5$ GeV/c.  The $f'_2(1525)$, a conventional $q\overline q$
meson is most prominent in (c), while the $f_J (1710)$, a suspected
glueball, is clearest in (a).  The $\pi^+\pi^-\pi^+\pi^-$ invariant
mass for (d)$dP_T<0.2$ GeV/c, (e)$0.2 {\rm GeV/c} dP_T<0.5$ GeV/c,
(f)$dP_T<0.5$ GeV/c.  The conventional $f_1(1285)$ is strong in (f),
but much weaker at lower $dP_T$, while the $f_0(1500)$ and $f_2(1930)$
are clearest at lower $dP_T$.  This figure is from Ref. 14.}
\label{fig:ptspectra}
\end{figure}

The difference can be seen by selecting events on the basis of $dp_T$,
the $p_T$ difference between the two momentum transfers.  Figure
\ref{fig:ptspectra} shows two example of the spectral changes in
different $dP_T$ bins.  Conventional mesons are prominent at high
$dP_T$, while at low $dP_T$, 'unconventional' mesons appear clearly.
Partial wave analyses were used to confirm the spin/parity
assignments.  Similar spectral changes have been observed for many
other channels, and the statistical significance of the effect is
unassailable.

Many theoretical interpretations have been given for this effect.  It
may be evidence that the Pomeron is a vector particle, or at least
acts as a non-conserved vector current,~\cite{vectors} in contrast to
the expected spin 0 behavior.  It has also been cited as evidence for
a QCD scale anomaly,~\cite{dima} and for the existence of
instantons.~\cite{instantons} The $p_T$ filter could also be due to a
large contribution from the gluon triangle diagram, Fig. 1b.
Alternately, it may be that Reggeon-Reggeon (meson-meson) and
meson-Pomeron interactions are still important at SPS energies.

RHIC should be able to shed much light on this effect.  If the $p_T$
filter effect is due to meson exchange, then it should be much reduced
at RHIC.  If the Pomeron acts as a vector particle, then it is likely
to display spin effects which should be visible in polarized
collisions.

The cross section for double-diffractive interactions at RHIC is about
100-200 $\mu$b,~\cite{doubledcross} corresponding to an interaction
rate of 20 kHz, even at a a center of mass energyy of 200 GeV and a
luminosity of $10^{31}$cm$^{-2}$s$^{-1}$.  RHIC will produce 100's of
millions of diffractive final states in a one month run, and any rate
limitation is likely to be in the detector or trigger.  The production
rates are very high even for rather specific final states.

\section{Hard $pp$ Diffraction}

Precision measurements of hard diffraction in $ep$ and $p\bar{p}$
scattering at the HERA and the Tevatron, respectively, raise many
questions about factorization and the universality of the hard
Pomeron.  Measurements of the diffractive structure function
$F_2^{D(3)}$~\cite{prn} at HERA and associated studies of the
diffractive-induced hadronic final state~\cite{difj} imply a universal
Pomeron composed predominantly of gluons. However, efforts to apply
the next-to-leading order QCD fits from $ep$ data to jet production in
$p\bar{p}$ scattering have not been entirely successful.

\begin{figure}
\center{\epsfig{figure=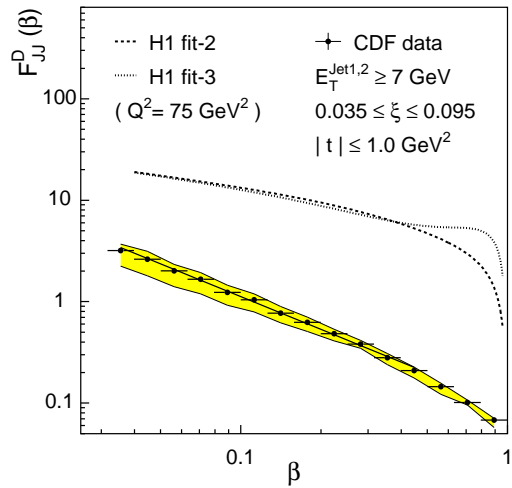,height=2.8in,clip=}}
\caption[]{The diffractive structure function as a function of
$\beta$, the parton momentum fraction in the Pomeron, derived from
dijet production at the Tevatron compared to next-to-leading order QCD
fits from $ep$ scattering at HERA. This figure is from
Ref. 21; further detail is given there.} 
\label{fig:cdffig}
\end{figure}

The structure of the Pomeron can be measured in terms of $\beta$, the
fraction of the total Pomeron momentum carried by the interacting
parton. The $\beta$ dependence of the diffractive structure function
$F^D_{JJ}$, as derived from jet measurements at CDF,~\cite{dino1} is
shown in figure~\ref{fig:cdffig} for a given region of $\xi$, the
fraction of the proton momentum carried by the Pomeron. A comparison
is made with parameterizations of the diffractive structure function
measured at HERA.  The $\beta$ dependence is similar although the
overall normalization is inconsistent between the two scattering
environments. Possible reasons for this include the need for a
modified Pomeron flux at Tevatron energies owing to unitarity
constraints or poorly understood gap-destroying multiple interactions
present in hadronic collisions.

The study of $pp$ scattering at RHIC can extend this field. With
center of mass energies up to several hundred GeV, RHIC covers a
kinematic region of diffractive scattering which overlaps with HERA
and the Tevatron. If Roman Pot detectors were installed around one of
the RHIC experiments then proton tagging would allow the study of the
$t-$dependence of the diffractive exchange.  A better understanding of
proton dissociation would help to reduce one of the dominant
systematic errors in diffraction measurements.

Further, measurements of single and double spin asymmetries will probe
the spin structure of the hard Pomeron from a polarized proton.  The
cross sections for most diffractive processes in $p\overline p$
collisions are usually a small percentage (typically around 1\%) of
the cross section of the corresponding non-diffractive final state.

Nothing is known about the spin sensitivity of this production.
Non-diffractive $W^\pm$, $Z^0$ and jet production comes largely from
$q\overline q$ annihilation and is expected to show significant
polarization asymmetries;~\cite{spin} diffraction might show similar
asymmetries. However, if diffraction is mediated by spin-0 Pomerons,
then, to lowest order, no asymmetry is expected.  At lower energies,
there is some evidence that diffractive reactions may be sensitive to
spin.~\cite{martin} Either case would be very interesting.

In addition to measurements sensitive to the hard scales of the
scattering such as jet production, the RHIC detectors can track
particles down to low $p_T$ in high-multiplicity events; these
abilities will allow for high-quality measurements of particle
multiplicities.  Studies of multiplicity fluctuations have been used
in the past to study the gluonic nature of the Pomeron.~\cite{edmul}

The rates for hard diffractive events may be scaled from the rates for
corresponding hard non-diffractive events.  Here, we give the rates
for a 1 month ($10^6$ s) $pp$ run at the maximum 500 GeV center of
mass energy, with an assumed luminosity of $10^{32}$cm$^{-2}$s$^{-1}$.
The D0 collaboration finds that the rate of diffractive to
non-diffractive jet events is
$1.07\pm0.10^{+0.25}_{-0.13}\%$.~\cite{dorate} Assuming that this
ratio holds, RHIC will produce about 2 million diffractive events per
year containing a jet with pseudorapidity $|\eta|<1$ and transverse
energy $E_T>30$ GeV.~\cite{rhicjets}

The CDF collaboration finds that diffractive $W^\pm$ production is
$1.15\pm0.55\% $ of non-diffractive $W^\pm$ production.~\cite{wrates}
With this ratio, each year RHIC will also produce 700 diffractive
$W^+\rightarrow e^+\nu$, 20 $W^-\rightarrow e^-$, and 4
$Z^0\rightarrow e^+e^-$, with lepton rapidity
$|\eta|<1$.~\cite{vbrates} The jet rates are very high, allowing for
precision studies of single and double spin asymmetries and the like.
Even diffractive vector boson production should be clearly visible.

\section{Diffraction with Heavy Ions}

The $pA$ program at RHIC enables the first studies of high energy 
diffraction with large and varying target sizes to take place. 

Diffraction in $pA$ collisions can be an important probe of the gluon
density of heavy nuclei at low-$x$, where gluon densities are expected
to be very high, and, even with shadowing, may saturate the available
phase space.~\cite{dumitru} In this regime, it may be possible to
describe the gluons as a classical field, and new effects, such as a
colored glass condensate may be observed.~\cite{glass}

Diffractive processes are sensitive to the total absorptive content of
the nucleus, and, through that, to the gluon content of the medium.
Several specific processes have been proposed to study the gluon
density at small $x$.  For example, one could compare the rate of
forward jet production in $pA$ to that in $pp$ to study nuclear
modifications of the gluon distribution.

The $t-$dependence of the diffractive exchange depends critically on
the target size. The $t$ and $A$ variation of diffractive cross
sections could be used to infer the relative contributions of soft and
hard diffraction and their variation with $A$.

The breakdown in universality of the picture of the partonic Pomeron
developed in deep-inelastic $ep$ scattering could be probed as a
function of $A$. Using $pp$ scattering as a baseline, $A-$dependent
deviations of multi-jet production and multiplicity fluctuations from
the simple partonic picture could be studied. This should be sensitive
to low-$x$ gluon shadowing and gap-destroying multiple interactions.

These studies complement the existing RHIC program for diffraction in
$AA$, in the form of ultra-peripheral photon-Pomeron and two-photon
interactions.~\cite{reviews} The cross sections for photon-Pomeron
fusion into vector mesons are very large,~\cite{vmprod} and $\rho^0$
production is observed at the predicted level.~\cite{STAR}
Ultra-peripheral collisions may be used to study nuclear shadowing;
production of both open heavy quarks~\cite{hq} and
quarkonium~\cite{strikman} is sensitive to the gluon distribution in
heavy nuclei.

\section{Conclusions}

With its polarized beams, high luminosity, and high-quality detectors,
RHIC is capable of making significant contributions to our
understanding of diffraction, in the areas of meson spectroscopy, hard
diffraction, and the low-$x$ gluon distributions in nuclei.  Much of
this physics depends on measurements of the outgoing scattered
protons.  The proton kinematics could be determined by adding Roman
pot systems upstream and downstream of a central detector at RHIC.  We
thanks Andrew Kirk for permission to use Fig. 2.  This work was
supported by the U.S. Department of Energy under Contract
No. DE-AC-03076SF00098.

\end{document}